\documentclass[prd,onecolumn,floatfix,preprintnumbers,showpacs,nofootinbib]{revtex4}
\usepackage{latexsym}
\usepackage{epsfig}

\date{\today}

\begin{document}

\title{Radion-Higgs mixing effects on bounds from LHC Higgs Searches}

\author{ Hiroshi de Sandes$^{a}$ and Rogerio Rosenfeld$^{b,c}$}
\affiliation{
$^a$Institut de Physique Th\'eorique, CEA-Saclay, 91191 Gif-sur-Yvette, France \\
$^b$CERN, Theory Division, CH-1211, Geneva 23, Switzerland \\
$^c$Instituto de F\'{i}sica Te\'orica, Universidade Estadual Paulista, 
Rua Dr. Bento T. Ferraz, 271, S\~ao Paulo, SP 01140-070, Brazil \\
}

\begin{abstract}
The radion, a scalar particle associated with the radius of a compact warped extra dimension, 
may be the lightest new particle in this class of models.   
Its couplings to SM particles are proportional to the their masses, similar to the usual Higgs boson,
but suppressed by a scale $\Lambda_r$, the radion vacuum expectation value. 
The main differences are the coupling to massless gauge bosons that receives contribution
from the trace anomaly of the energy-momentum tensor due to the the nonvanishing $\beta$ functions and
the mixing with the Higgs boson arising from a nonminimal coupling to gravity
parametrized by a dimensionless coefficient $\xi$. 
In particular, these differences can result in significant modifications of the radion phenomenology.
We use current LHC data on Higgs searches to find exclusion regions on 
the parameters of a radion model, $\Lambda_r$, $\xi$ and the radion mass $m_r$.
We find that, even for low values of $\Lambda_r$,
the radion can still have a mass in the region where the Standard Model Higgs has been
excluded, for a narrow range of values for the mixing parameter. Some signals at the LHC
for this scenario are discussed.
We also find that it is possible to hide the Higgs boson in the current searches
in this model, due to a suppression of its couplings.
\end{abstract}

\preprint{CERN-PH-TH/2011-316}
\maketitle

\section{Introduction} 
One of the leading candidates to
solve some of the well-known problems of the Standard Model (SM), 
such as the hierarchy problem,
is a class of models with warped extra dimensions \cite{rs}.
These models are characterized by the existence of a compact, curved 
extra dimension with AdS geometry containing 
two 4-dimensional boundaries in the extra dimension, the so-called branes.

In these models the otherwise unnatural difference between the Higgs mass and the
Planck scale is explained by an exponential warp factor between the two branes
arising from the AdS geometry, which fixes the ratio of the size of the extra 
dimension to the AdS curvature radius, with the additional assumption
that the Higgs is localized in one of the branes (the TeV brane).
The fermion mass hierarchy arises naturally by ${\cal O}(1)$ parameters that control the fermion
wave functions in the extra dimensions, which in turn determines the Yukawa couplings to 
the TeV brane localized Higgs boson \cite{huber}. The same fermion wave functions also
govern their potentially flavor-changing neutral current couplings to Kaluza-Klein (KK) excitations of the gauge bosons,
and as a result they are strongly suppressed for light quarks. This constitutes the so-called RS-GIM mechanism \cite{RSGIM},
although some tension still exists in the KK contribution to  $K - \bar{K}$ oscillations. \cite{tension}. 

As in all models of extra dimensions, warped models predict the existence of
Kaluza-Klein excitations of the SM particles with mass scales determined by the compactification
scale. In order to be consistent with electroweak precision tests, this scale is usually in the
few TeV range. Detailed studies of the production and detectability of these excitations have been
performed in the literature \cite{pheno}.

Another interesting consequence of these models arises from the fact the presence of the two branes
breaks scale invariance, leading to the the existence of a dilaton-like particle, usually
called radion. The radion, which we will denote by $r$, has an important role in stabilizing 
the size of the extra dimension \cite{gw}.
As a dilaton, the radion couples to the trace of 
the energy-momentum tensor of the theory.
Hence, its couplings are proportional to masses of particles, in much the same way as the usual Higgs boson.
As opposed to the KK excitations of SM particles, the radion mass is not determined
by the compactification scale. It depends on the details of the mechanisms that stabilizes the 
size of the extra dimension and hence can be considered as a free parameter.
It might very well be the lightest new particle in these models.
Many phenomenological studies have been dedicated to finding signals for the radion
at colliders \cite{grw,cgk,cdhy,dggt,bcrdg,chl,leshouches,elps}.
Some studies were also performed by ATLAS and CMS 
a few years ago to assess the LHC reach for a radion decaying to $\gamma \gamma$, $Z Z^\ast$ and
$hh$ \cite{atlas,cms}.
In fact, many of the Higgs searches can also be used for limiting the parameters of a  radion model \cite{bcrdg}.
This was already done using data from Tevatron \cite{tevatron}.
With the recent results from the LHC, it is
finally possible to put bounds on parameters of the radion.


One general feature of radion models is that the radion can in principle mix with the
Higgs boson. This mixing can have important phenomenological consequences.
If one does not include the mixing, the only way to suppress the radion couplings is to increase its energy scale.
Therefore, LHC bounds will result in large values for the radion scale \cite{fkt,barger} 
\footnote{Constraints on similar dilaton models without mixing were obtained in \cite{dilaton}}. 
However, including the mixing opens up a rich interplay with 
Higgs physics and also allows the possibility to evade the LHC Higgs exclusion constraints on the radion and
the mixed Higgs without increasing the new physics scale.

In this work we concentrate on the radion mediated processes 
$gg \rightarrow r \rightarrow \gamma \gamma$ and
$gg \rightarrow r \rightarrow V V^\ast $, where $V=W,Z$, and use
LHC data for Higgs searches up to a scalar mass of $600$ GeV in order to exclude
regions in the parameter space of the radion model.

The Higgs and radion cross sections can be easily related as follows. 
The proton-proton cross section for the production of a given final state $X$ via the radion in 
gluon-gluon fusion, which is the dominant process at the LHC with $7$ TeV center-of-mass energy, 
can be written as
\begin{equation}
\sigma(pp \stackrel{gg}
     {\rightarrow} r \rightarrow X) (s) = \int d\tau {\cal L}_{gg} (\tau) \hat{\sigma}(\tau s)
\end{equation}
where ${\cal L}_{gg} (\tau)$ is the gluon 
luminosity given by the parton distribution function 
of the gluons in a proton and $\hat{\sigma}$ is the partonic cross section. 
Using a Breit-Wigner narrow width approximation one can relate the SM Higgs
cross section to a radion cross section with the same mass by
\begin{equation}
\frac{\sigma(pp \stackrel{gg}
     {\rightarrow} r \rightarrow X)}{\sigma(pp \stackrel{gg}
     {\rightarrow} h \rightarrow X)_{SM}} = \frac{\Gamma_h^{SM}}{\Gamma_r} \frac{\Gamma(r\rightarrow gg)}{\Gamma(h\rightarrow gg)_{SM}} 
\frac{\Gamma(r\rightarrow X)}{\Gamma(h\rightarrow X)_{SM}}
\label{enhancement}
\end{equation} 
where $\Gamma_h^{SM}$ and $\Gamma_r$ are the total SM Higgs and radion widths respectively.
This should be a very good approximation since for most of the ranges of masses we will be considering
both the Higgs boson and the radion are very narrow resonances ($\Gamma/M <<1$).
In the next section we describe the model we will adopt to calculate the bounds on the radion.

\section{The model and radion couplings}

In this work we follow closely the realistic warped space scenario for the radion described 
by Cs\'{a}ki et al. \cite{chl}.
The unperturbed metric is written as:
\begin{equation}
ds^2 = \left( \frac{R}{z} \right)^2 \left( \eta_{\mu\nu} dx^\mu dx^\nu - dz^2 \right),
\end{equation}
where $z$ refers to the coordinate in the 5th dimension restricted to $R<z<R'$, and $k = 1/R$ is
the AdS curvature. The radion is related to the scalar perturbation of the metric, which at leading order is
given by:
\begin{equation}
\delta g_{M N} = -2 F \left( \frac{R}{z} \right)^2 \left( \begin{array}{cc}
\eta_{\mu\nu} & 0  \\
0 & 2   \end{array} \right)
\end{equation}
where $F(x,z)$ is the 5d radion field.

The linear radion couplings are determined by the modification
of the action due to the linear perturbation of the metric, which by the definition of the 
energy-momentum tensor $T^{M N}$ is given by:
\begin{equation}
\delta S = -\frac{1}{2} \int d^5 x \sqrt{g} T^{M N} \delta g_{M N} = 
 \int d^5 x \sqrt{g} F ( \mbox{Tr} T^{M N} - 3 g_{55} T^{55})
\end{equation}

The canonically normalized scalar radion field in 4d is related
to $F(x,z)$ by:
\begin{equation}
r(x) = \Lambda_r \left( \frac{R'}{z} \right)^2 F(x,z)
\end{equation}
where $\Lambda_r = \sqrt{6}/R'$ can be interpreted as the radion vacuum expectation value that leads
to the stabilization of the size of the extra dimension.
For fields that are strongly localized in the infrared brane, such as the Higgs boson and the top quark,
the coupling to the radion is given by
the usual term 
\begin{equation}
{\cal L} = \frac{r(x)}{\Lambda_r} T^{\mu}_{\mu}
\end{equation}

The leading contribution to the radion interaction with massive gauge bosons $W^\pm$ and $Z$ and  
heavy fermions localized in the IR brane 
(there are corrections depending on the localization parameters $c_L$ and $c_R$ that are small when the 
fermions are localized in IR such as the top quark and they are not large for the $b$ quark, see \cite{chl})
is similar to the Randall-Sundrum set-up
\begin{equation}
{\cal L}_{r(VV,\bar{\psi} \psi)} = -\frac{r}{\Lambda_r} 
 \left( 2 M_W^2 W^2 + M_Z^2 Z^2 - m \bar{\psi} \psi  \right).
\end{equation}

Usually the coupling of the radion to massless gauge bosons vanishes at tree level. At 1-loop level 
it arises due to two
contributions: the trace anomaly, which is related to the beta function, and the top quark triangle diagram.
However, in the warped scenario, there are two main differences: a tree level bulk contribution from radion and gauge
bosons wave functions and a modification in the beta function term to take into account that only particles in the 
infrared brane contribute to the running. However, when a proper matching at 1-loop between the 5d and 4d coupling constants is performed, the usual beta function is recovered. 
The final result for this coupling is \cite{chl}:
\begin{eqnarray}
{\cal L}_{rAA} &=& - \frac{r}{\Lambda_r} \left[
 \frac{\alpha}{8 \pi} \left( (b_2+b_Y) -  F_1(\tau_w) -  \frac{4}{3} F_{1/2}(\tau_t) \right)
 + \right. \\ \nonumber
 && \frac{1}{4\ln (R'/R)} \left( 1- 4 \pi \alpha_s \left(\tau_{UV}^{(0)}+\tau_{IR}^{(0)} \right) \right] 
F_{\mu \nu}  F^{\mu \nu}
\end{eqnarray}
for photons and 
\begin{eqnarray}
{\cal L}_{rgg} &=& - \frac{r}{\Lambda_r} \left[ \frac{\alpha_s}{8 \pi} \left(b_3 - \frac{1}{2} F_{1/2}(\tau_t) \right)
+ \right. \\ \nonumber
&& \frac{1}{4\ln (R'/R)} \left( 1- 4 \pi \alpha_s \left(\tau_{UV}^{(0)}+\tau_{IR}^{(0)} \right) \right] 
G_{\mu \nu}^a  G^{\mu \nu}_a
\end{eqnarray}
for gluons, where
$\tau_x = 4 m_x^2/m_r^2$ and the functions $F_{1/2}(\tau) \rightarrow (0,-4/3)$  and $F_1(\tau) \rightarrow (2,7)$  as $\tau \rightarrow (0,\infty)$ are defined in the usual way (see, e.g., \cite{grw}).  
The coefficients of the SM $\beta$ functions we use are 
$b_3 = 7$, $b_2 = 19/6$ and $b_Y = −41/6$. 
The parameters
$\tau_{UV}^{(0)}$ and $\tau_{IR}^{(0)}$ are related to the Planck and TeV-brane induced kinetic
terms. The first term in the brackets is the usual radion contribution due to the QCD trace anomaly 
and the top triangle contributions. The other terms arise from the integration of the radion and gauge fields profiles in the bulk and the inclusion of brane-localized gauge field kinetic terms.
It is interesting to notice that the term arising from the integration of the profiles resulting in the $1/[4\ln (R'/R)]$ is actually of the same order of $\alpha_s/8 \pi$ for a realistic value of $\ln (R'/R) = 30$  and hence should not be neglected. 

For the Higgs boson the situation is more involved because of two factors:
spontaneous symmetry breaking and the fact that the
energy-momentum tensor of a scalar field must be modified in order to allow the possibility that its trace can
vanish in the zero-mass limit in the case of $\xi = 1/6$, resulting in a classically conformal invariant theory \cite{ccj}.

For a Higgs lagrangian (after symmetry breaking)
\begin{equation}
{\cal L}_h = \frac{1}{2} (\partial_\mu h)^2 - \lambda \left( \frac{(h+v)^2}{2} - \frac{v^2}{2} \right)^2
\end{equation}
the modified energy-momentum tensor $\Theta_{\mu \nu}$ reads \cite{ccj} :
\begin{equation}
\Theta_{\mu \nu} = \partial_\mu h \partial_\nu h - \eta_{\mu \nu} {\cal L}_h + 
\xi \left(  \eta_{\mu \nu} \partial_\lambda \partial^\lambda - \partial_\mu \partial_\nu \right) 
\left( \frac{(h+v)^2}{2} \right)
\end{equation}
which leads to
\begin{equation}
\Theta^{\mu}_\mu = -(1-6\xi) (\partial_\mu h)^2 +  (1-6\xi) (\lambda h^4 + 4 \lambda v h^3) + (4-30 \xi) \lambda v^2 h^2
- 12 \xi \lambda v^3 h.
\end{equation}
Therefore, for $\xi = 1/6$, one gets
\begin{equation}
\Theta^{\mu}_\mu = - \lambda v^2 h^2 - \frac{1}{2} \lambda v^3 h
\end{equation}
where the first term of the trace of the modified energy-momentum tensor is proportional to the
Higgs mass whereas the second term will induce a mixing between the radion and the Higgs boson.

Radion phenomenology is very sensitive to the values of $\xi$. The $\xi$ term for a general scalar field $\phi$
can be written as a coupling to the Ricci scalar R as
\begin{equation}
{\cal L}_\xi = \xi R \phi^2
\end{equation}
and it breaks a shift symmetry in the scalar field. In models where the Higgs is a Goldstone boson,
one would expect the residual shift symmetry to forbid such a term, which correspondes to setting $\xi = 0$ \cite{grw}.
Even if the Higgs is an approximate Goldstone boson $\xi$ should be small.
However, in general one can consider $\xi$ as a free parameter. 
A non-zero value of $\xi$ results in radion-Higgs mixing, which can be important for 
phenomenological purposes such as the goal of this letter.

The radion-Higgs mixing can be studied following the results of \cite{grw,cgk,cdhy,dggt}. One can write
the physical radion and Higgs fields that diagonalize the quadratic terms and have canonical normalization
as:
\begin{eqnarray}
r &=& a \; r^{{\mbox{\tiny phys}}} + b \; h^{{\mbox{\tiny phys}}} \\ \nonumber
h &=& c \; r^{{\mbox{\tiny phys}}} + d \; h^{{\mbox{\tiny phys}}}
\end{eqnarray}
The physical masses $m_r$ and $m_h$ and the parameters $\Lambda_r$ and $\xi$
determine the mixing parameters $a,b,c$ and $d$. Their explicit expression can be found
in, {\it e.g.}, Dominici {\it et al.} \cite{dggt}.
In the case of no mixing ($\xi = 0$), $a=-d=-1$ and $b=c=0$.

\section{Phenomenology}

We will be interested in the case where the radion is heavier than the Higgs boson.
It will be considered
the cases where the Higgs boson is outside the region where the
Standard Model Higgs has been recently excluded at $95 \%$ CL, $m_h<131$ GeV \cite{CernSeminars}. We will also comment on the 
possibility of hiding a heavier Higgs bosons.
We denote in the following the ratio of the
SM vacuum expectation value to the radion scale  $\gamma = v/\Lambda_r$.
The effects of the radion potential arising from the stabilization mechanism, such as 
triple and quartic radion couplings, and
brane-localized kinetic terms are not included, since both effects are very model-dependent.
  
The ratio of the radion width to the same-mass SM Higgs boson width is 
the same in the fermionic and electroweak gauge boson channels ($V=W,Z$) (neglecting again small contributions 
from the fermion localization factors)
and is given by \cite{dggt}
\begin{equation} \label{eq:ratiovv}
\frac{\Gamma(r\rightarrow V V^\ast, \bar{f}f)}{\Gamma(h\rightarrow V V^\ast,\bar{f}f)_{SM}} = 
(c + \gamma a)^2 = 
\left( -1 + 3 \xi + 3 \xi \frac{m_r^2+m_h^2}{m_r^2-m_h^2} \right) \gamma^2 + {\cal O}(\gamma^3).
\end{equation}
Since these decays modes are dominant for most regions of parameter space, the branching ratio of the radion and the Higgs in these channels are similar. The exception occurs when there is a cancellation such that 
$(c + \gamma a) \approx 0$, in which case the radion becomes phobic to both fermions and vector bosons.
In this case there will be much weaker bounds from the LHC, since the decay rates will be decreased, 
but the production rate remains significant since it is dominated by the anomaly. 
This region of parameter space will show up in our results. 


The ratio of the radion width into gluons to the width of a same mass SM Higgs boson is
\begin{equation}
\frac{\Gamma(r\rightarrow gg)}{\Gamma(h\rightarrow gg)_{SM}} = 
\left| \frac{ (c+ \gamma a)   F_{1/2}(\tau_t) - \gamma a \left( 2 b_3 + \frac{16 \pi /\alpha_s}{ 4 \ln R'/R} \right)}{ F_{1/2}(\tau_t)} \right|^2,
\end{equation}
and similarly for the photon channel
\begin{equation}
\frac{\Gamma(r\rightarrow \gamma \gamma)}{\Gamma(h\rightarrow \gamma \gamma)_{SM}} = 
\left| \frac{ (c+ \gamma a) \left(  F_1(\tau_w) +\frac{4}{3}F_{1/2}(\tau_t) \right) - \gamma a \left(  (b_2+b_Y) + \frac{8 \pi/\alpha}{ 4 \ln R'/R} \right) 
 }{ F_1(\tau_w) +\frac{4}{3}F_{1/2}(\tau_t) } \right|^2.
\end{equation}

When kinematically allowed, the radion can also decay into 2 Higgs bosons.
We use the Feynman rules given in \ref{dggt} to derive 
\begin{eqnarray}
\Gamma(r\rightarrow hh) &=& \frac{\gamma ^2}{32 \pi  v^2 m_r} \left|C+2 A m_h^2+B m_r^2\right|^2 \sqrt{1-\frac{4 m_h^2}{m_r^2}} \\
A &=& 6 b \xi  (\gamma  (a d+b c)+c d)+a d^2 \nonumber \\
B &=& d (12 a b \gamma  \xi +a d (6 \xi -1)+2 b c) \nonumber \\
C &=&\left(-4 d (a d+2 b c)-\frac{3 c d^2}{\gamma }\right){m_{h_{0}}}^2 \nonumber
\end{eqnarray}

Finally, for the partial and total widths of the SM Higgs boson we use the results from 
HDECAY code \cite{hdecay} and write for the total radion width:
\begin{eqnarray}
\Gamma_r &=&  \frac{\Gamma(r\rightarrow V V^\ast, \bar{f}f)}{\Gamma(h\rightarrow V V^\ast,\bar{f}f)_{SM}} \left(\Gamma(h\rightarrow V V^\ast)_{SM} + 
\sum_{\tau,b,t,c}\Gamma(h\rightarrow \bar{f}f)_{SM} \right)  \\ \nonumber
&& + \frac{\Gamma(r\rightarrow gg)}{\Gamma(h\rightarrow gg)_{SM}} \Gamma(h\rightarrow gg)_{SM} 
+ \frac{\Gamma(r\rightarrow \gamma\gamma)}{\Gamma(h\rightarrow \gamma\gamma)_{SM}} \Gamma(h\rightarrow \gamma\gamma)_{SM} +\Gamma(r \rightarrow hh) 
\end{eqnarray}

\section{Results}

In what follows we take as free parameters of the radion model
the physical radion mass, $m_r$, the radion vacuum expectation value, $\Lambda_r$ and the mixing parameter
$\xi$. As mentioned before we will consider the case where the mixed Higgs boson mass is $m_h < 131$ GeV, 
so that it has escaped detection so far and also the case where it is inside the region where the
Standard Model Higgs has been excluded. We will fix $\ln R'/R = 30$ in 
order to solve the large hierarchy problem.

One must notice that for a given value of $m_r$ and $\Lambda_r$ there is a limited range of values that the mixing parameter 
can assume, given by the condition that the unmixed masses are real.
There is another bound on $\xi$ arising from perturbative unitarity 
in $WW$ scattering \cite{hkm}. However, requiring that perturbative unitarity is maintained 
up to a scale of ${\cal O}$(TeV) results in the two bounds being similar.
We take into account this maximum range of $\xi$ in our results.

We use the CMS data on the Higgs search with approximately $5$ fb$^{-1}$ integrated luminosity
recently presented at CERN \cite{CernSeminars}, considering
only the individual channels $\gamma \gamma$, $WW \rightarrow 2l2\nu$, $ZZ \rightarrow 4l, 2l2\nu$. 
The so-called CL$_s$ method is used
to put bounds at 95\% confidence level on  $ \mu = \sigma/\sigma_{SM}$, 
the signal strength modifier, where $\sigma$ is the ``true" cross section and
$\sigma_{SM}$ is the predicted SM cross section. It is assumed the SM branching ratios 
throughout their analysis. Due to the similarities between the radion and the Higgs boson, 
it is straightforward to use the data on Higgs searches to constrain parameters in the radion model.
In this case the signal strength modifier is simply given by eq.(\ref{enhancement}), where the 
corrections for the different branching ratios are taken into account. 
Notice that because there are always ratios of quantites for the Higgs boson and for the radion of the same mass, 
one expects that higher order corrections such as NLO QCD K-factors will be cancelled, 
since their production mechanism is the same. 


 
\begin{figure}[htb]
\begin{center}
\includegraphics[width=7cm]{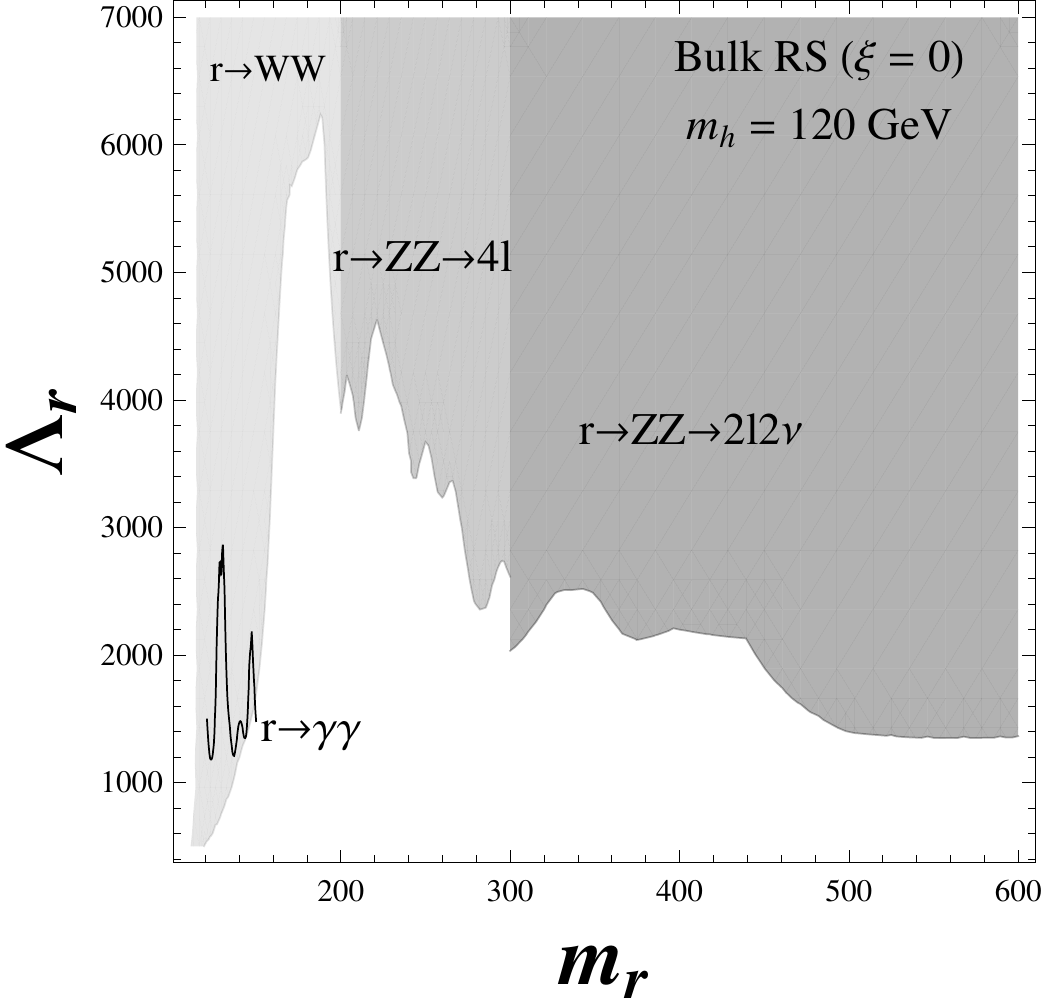}
\end{center}
\caption { Bounds on $\Lambda_r$ and $m_r$ coming from CMS data for a pure radion ($\xi=0$) and a light higgs 
$m_h=120$ GeV. The white regions are excluded whereas the colored regions are allowed. The light gray region is the one allowed
by data from $WW$ channel and the dark gray from $ZZ$ channel. The region above the solid line is allowed by data from $\gamma\gamma$ 
channel.}
\label{Exclusionxizero}
\end{figure}

In Fig.(\ref{Exclusionxizero}) we show the exclusion regions for a pure radion with no-mixing  ($\xi = 0$)
for the different channels, with a Higgs mass fixed at  $m_h=120$ GeV.
One can see that there are regions where the bounds are very stringent.
For instance, for $\Lambda_r = 1.8$ TeV, a radion mass in the range $160< m_r < 440$ GeV is already excluded.
It is only possible to have a pure radion with a mass $m_r=180$ GeV if its
couplings are suppressed by $\Lambda_r > 6$ TeV. However, for $m_r=280$ GeV the bounds are less stringent, allowing for 
$\Lambda_r = {\cal O}(3)$ TeV.
  
\begin{figure}[htb]
\begin{center}
\includegraphics[width=6.5cm]{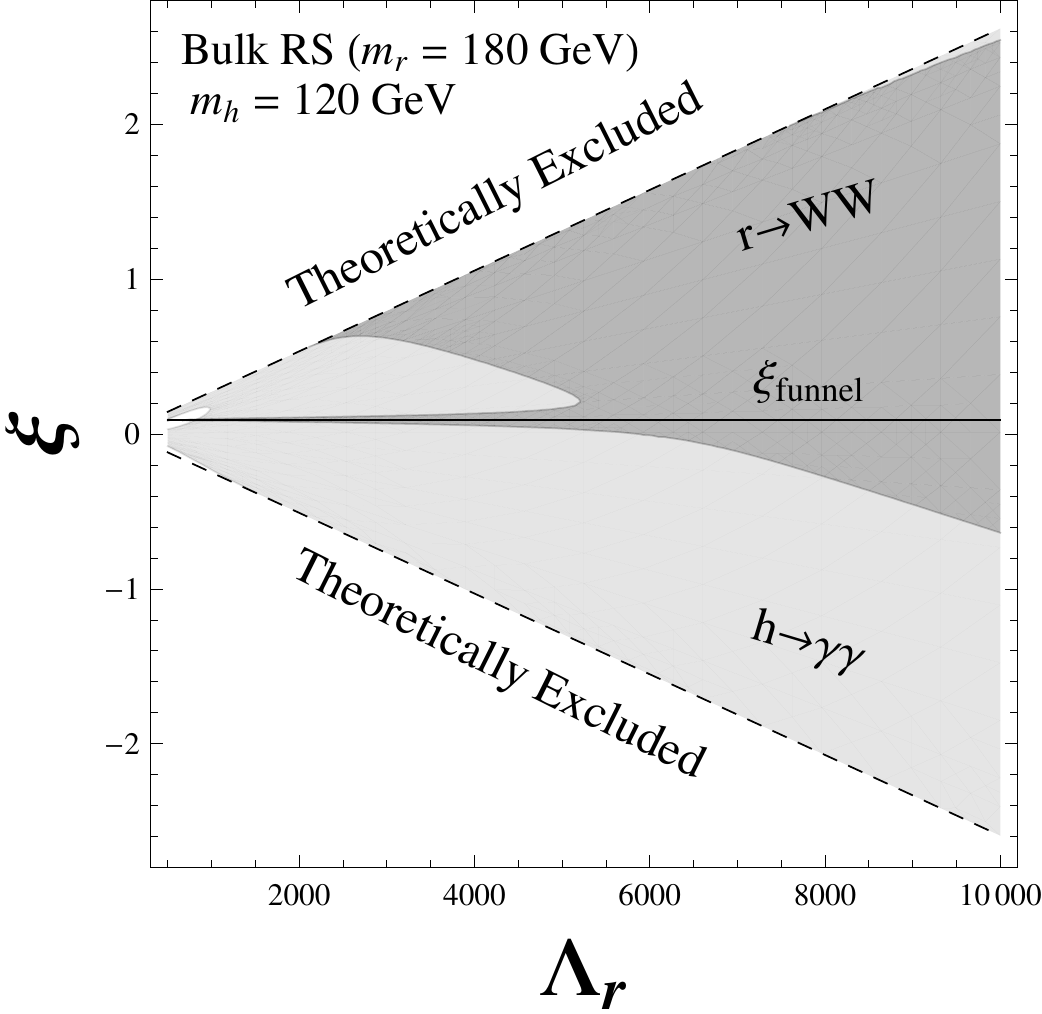}
\includegraphics[width=6.5cm]{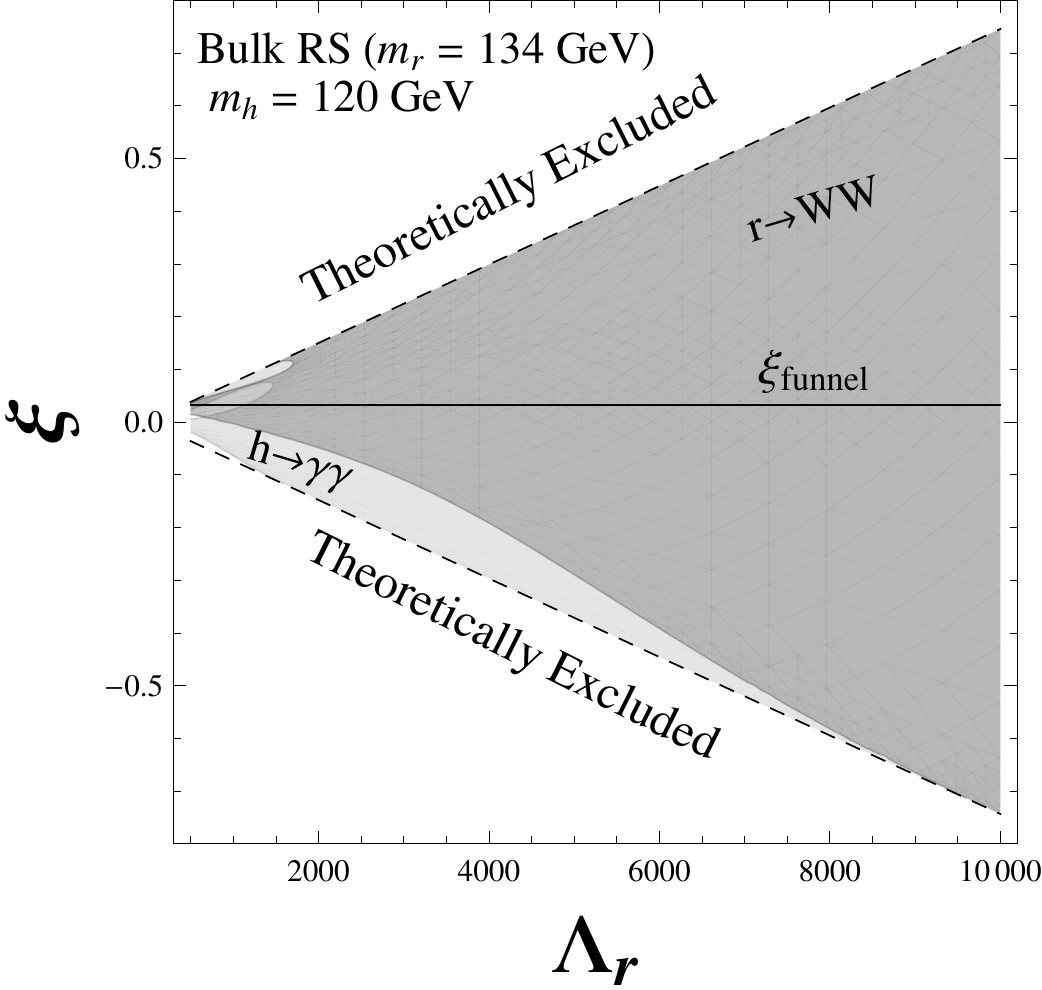}
\end{center}
\caption{Bounds on $\Lambda_r$ and $\xi$ coming from CMS combined for two representative values of 
$m_h = 120$ GeV and $m_r = 134, 180$ GeV, corresponding to a loosely constrained and tightly constrained situation respectively.
The white regions are excluded whereas the colored regions are allowed. 
The dotted lines correspond to the minimum and maximum 
values of $\xi$ theoretically allowed for a given value of $\Lambda_r$, $m_h$ and $m_r$. 
The region above (under) the maximum (minimum) values of $\xi$ is theoretically excluded. The white regions inside the
theoretically allowed $\xi$ range are experimentally excluded.
The light gray region is the one allowed by data for the Higgs and the dark gray for the radion. The intersections
of these two regions are the darker ones. The solid line are the values for $\xi_{funnel}$.
}
\label{Exclusionlighthiggs}
\end{figure}

The LHC constraints can be ameliorated with the inclusion of a non-zero mixing parameter. 
In Fig.(\ref{Exclusionlighthiggs}) we show some of the exclusion regions in the $(\xi,\Lambda_r)$ plane from the CMS Higgs search 
for two extreme scenarios where the $WW$ constraints are strong ($m_r=180$ GeV) and  
weak ($m_r=134$ GeV), fixing again the Higgs mass at $m_h=120$ GeV. 
We can see from the most constrained situation that a radion mixed with the Higgs is allowed all the way down to $\Lambda_r=1$ TeV, 
insofar as direct searches at the LHC are concerned.
This is possible because of the ``funnel" region for $\xi$ where the cancellation in the $VV$ couplings of 
the radion mentioned earlier occurs. The solid line that runs exactly in the ``funnel" are the values where the second order term in 
the Eq. (\ref{eq:ratiovv}) vanishes. From now on we will call these values $\xi_{funnel}$.
In the scenario 
with $m_h=120$ GeV and $m_r=134$ GeV, where the LHC constraints are weaker, there is a larger allowed region
of parameter space. We also show in the plots that, as expected, there is basically no constrains on the Higgs for this particular case.

\begin{figure}[h,t,b]
\centering
\includegraphics[width=6.3cm]{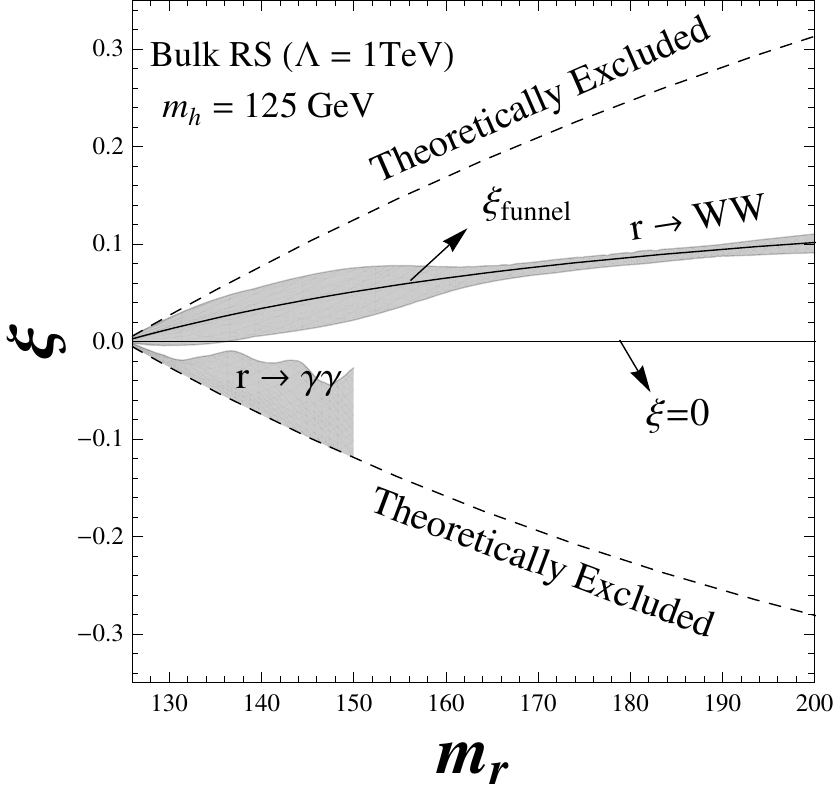}
\includegraphics[width=6.3cm]{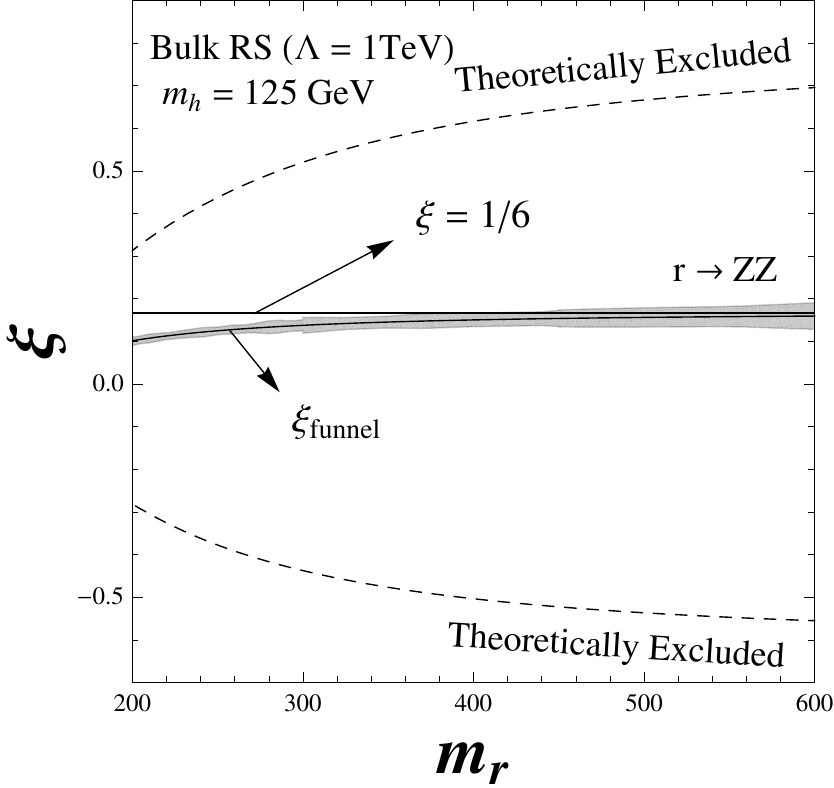}
\caption{Bounds on $\xi$ and $m_r$ coming from CMS data for $m_h = 125$ GeV for $\Lambda_r=1$ TeV.
The dotted lines correspond to the minimum and maximum 
values of $\xi$ theoretically allowed for a given value of $\Lambda_r$, $m_h$ and $m_r$. 
The region above (under) the maximum (minimum) values of $\xi$ is theoretically excluded. 
The dark gray regions are allowed by data from $\gamma\gamma$ and $WW$/$ZZ$ channel, both for the radion.
The intersections of these two regions are the darker ones. The light gray region is allowed by data from $\gamma\gamma$ for
the Higgs. The solid lines are the values for $\xi_{funnel}$, $\xi=0$ and $\xi=1/6$ as indicated by the arrows.
}
\label{ximr}
\end{figure}

In order to discuss in more detail these results we show in the Fig.(\ref{ximr}) 
the constraints in the ($\xi$, $m_{r}$) plane  
for $\Lambda_r=1$ TeV for two ranges of the radion mass sensitive to different search channels.
For $m_r < 200$ GeV (left figure), we show the $r \rightarrow \gamma \gamma$ and $r\rightarrow WW \rightarrow l \nu l \nu$
channels. For $m_r> 200$ GeV (right figure), we use the constraints arising from $r\rightarrow ZZ \rightarrow 2 l 2 \nu$ 
for $300<m_r<450$ GeV and
$r\rightarrow ZZ \rightarrow 4 l$ in the remaining range.
We fix $m_h=125$ GeV, which is favored by current LHC data.
One can see that $m_r < 150$ GeV is excluded in this case.
This limit is of course relaxed for larger values of $\Lambda_r$. 
For example, for $\Lambda_r = 2$, we verified that a radion with $m_r=140$ GeV
and a $\xi$ close to zero is allowed. In that case the radion decay channels are similar to the
SM Higgs but suppressed by the $\gamma$ factor so that the radion could appear in the 
$ZZ^{*}\rightarrow4l$ channel\cite{atlas}.

\begin{figure}[h,t,b]
\centering
\includegraphics[width=8cm]{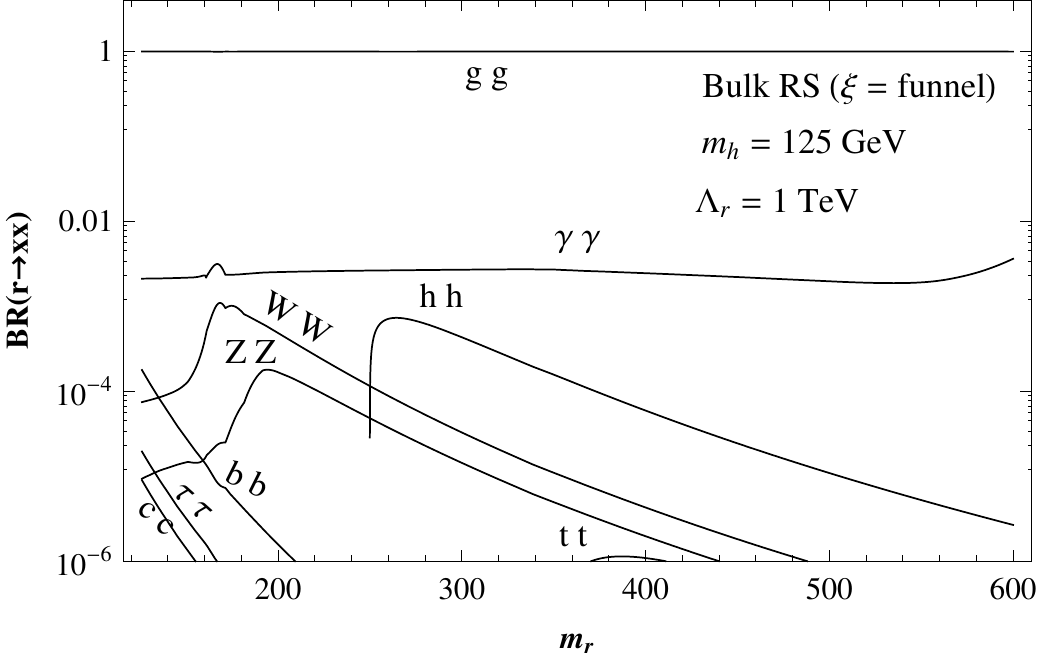}
\includegraphics[width=8cm]{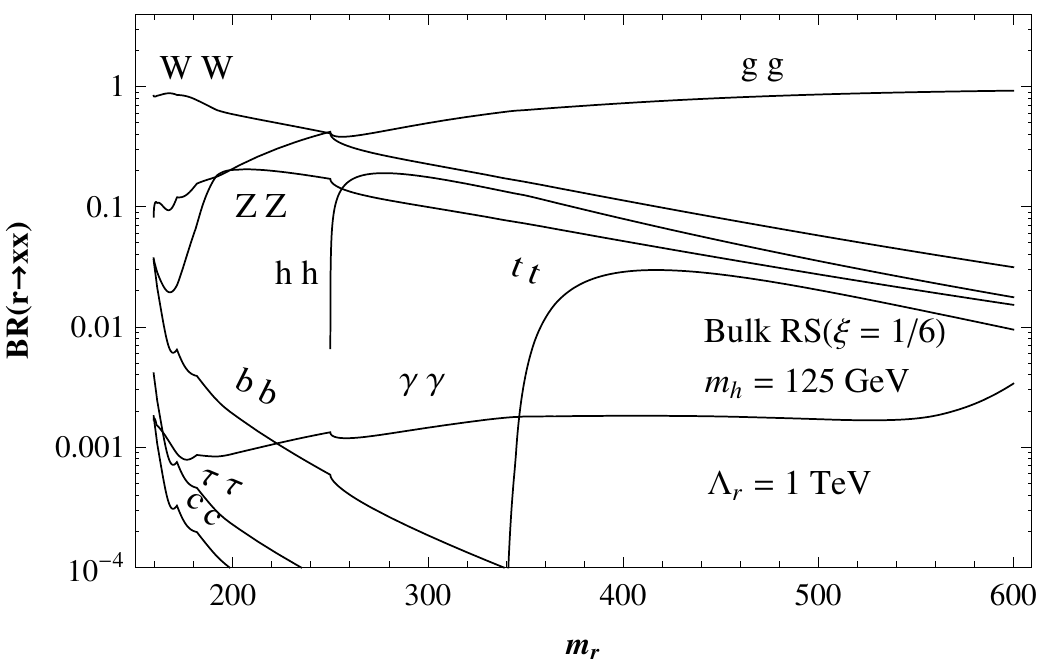}
\caption{Radion branching ratios for $\xi=\xi_{funnel}$(left) and $\xi=1/6$(right). We
fixed $m_r=125$ GeV and $\Lambda_r=1$ TeV.}
\label{brsxi}
\end{figure}

For $\Lambda_r = 1$ TeV, there is an allowed region for $m_r>150$ GeV 
constrained by the $WW$ and $ZZ$ channels to be near $\xi=\xi_{funnel}$. 
As was already explained, in the funnel region the couplings of the radion to fermions and 
massive gauge bosons are suppressed. One could expect that because the radion becomes
fermiophobic (and gaugephobic) its production cross section by gluon fusion will also get suppressed.
However, the gluon fusion mechanism will still be effective due to contribution from the trace anomaly.
This is shown in Fig.(\ref{brsxi}), where the radion branching ratios
as a function of its mass is given for $m_h=125$ GeV, $\Lambda_r=1$ TeV and two cases for $\xi$, $\xi=\xi_{funnel}$ (left figure)
and $\xi = 1/6$ (right figure).
In the $\xi=\xi_{funnel}$ case the radion could be only seen in the $\gamma \gamma$ channel.
The radion couplings are very sensitive to small variations in $\xi$. For the
region where the conformal value $\xi = 1/6$ is allowed by the LHC data, which happens for large values of the radion mass, 
one can have a significant $hh$ branching ratio. One can also see that the discovery process for a radion for $\xi = 1/6$ could be its decay to Higgs pairs.
Realistic studies of this possibility were already performed by ATLAS\cite{atlas} and CMS\cite{cms}. 
The most promising signal was found to be
the practically background free $gg\rightarrow hh \rightarrow \gamma\gamma \bar{b}b$, where a radion
mass reconstruction could be possible.

We also verified that if $\xi=\xi_{funnel}$, there could be an impact of the radion scenario in the low-mass Higgs searches
from the $\gamma\gamma$ channel, where there is an enhancement of at most $50 \%$ due to the mixing with the radion.
This is consistent with the recent results on the Higgs searches using almost $5fb^{-1}$ of data collected on 2011.
Both ATLAS and CMS \cite{CernSeminars} are observing an excess on the $H\rightarrow\gamma\gamma$ channel respectively for $m_{h}$ near $126$ and $124$ GeV
with a local significance of $3.6\sigma$ and $3.4\sigma$. Their best fit for the $H$ to $\gamma\gamma$ cross-section lay within 
the $1\sigma$ interval of $1-2.5$ times the SM Higgs cross-section that is consistent with our result of $1.5$ $\sigma_{SM}$ but
also with the SM Higgs hypothesis. Hopefully this scenario can be probed in the next years.

\begin{figure}[htb]
\begin{center}
\includegraphics[width=10.cm]{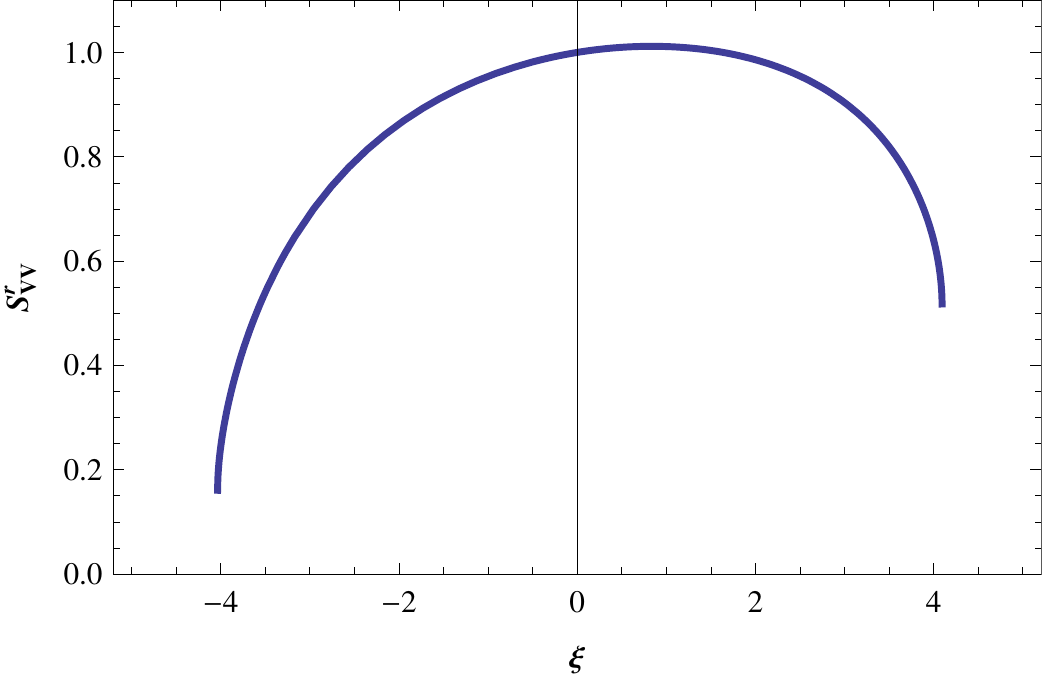}
\end{center}
\caption{Suppression of the Higgs production cross section in gluon-gluon fusion in the $VV$ channel
as a function of the mixing parameter $\xi$ for $m_r = 600$ GeV, $m_h = 300$ GeV and $\Lambda_r = 10$ TeV.}
\label{suppression}
\end{figure}

Finally, we study the interesting possibility to hide the Higgs boson in the context of the radion model.
Again, the Higgs-radion mixing is a crucial ingredient. The relevant factor in this case is the quantity we will denote by
$S^r_{VV}$, indicating the suppression of the Higgs signal in the radion model with respect to the SM model 
in the gluon fusion production mechanism considering the $VV$ decay channel
\begin{equation}
S^r_{VV} = \frac{\Gamma^r(H\rightarrow gg) \Gamma^r(H\rightarrow VV) / \Gamma^r(H)}
{ \Gamma^{SM}(H\rightarrow gg) \Gamma^{SM}(H\rightarrow VV) / \Gamma^{SM}(H)}.
\end{equation}
This suppresion is shown in Fig.(\ref{suppression}), where one can see a significant
reduction in the cross section for these values of the mixing parameter near the theoretical limits.

\begin{figure}[htb]
\begin{center}
\includegraphics[width=6.5cm]{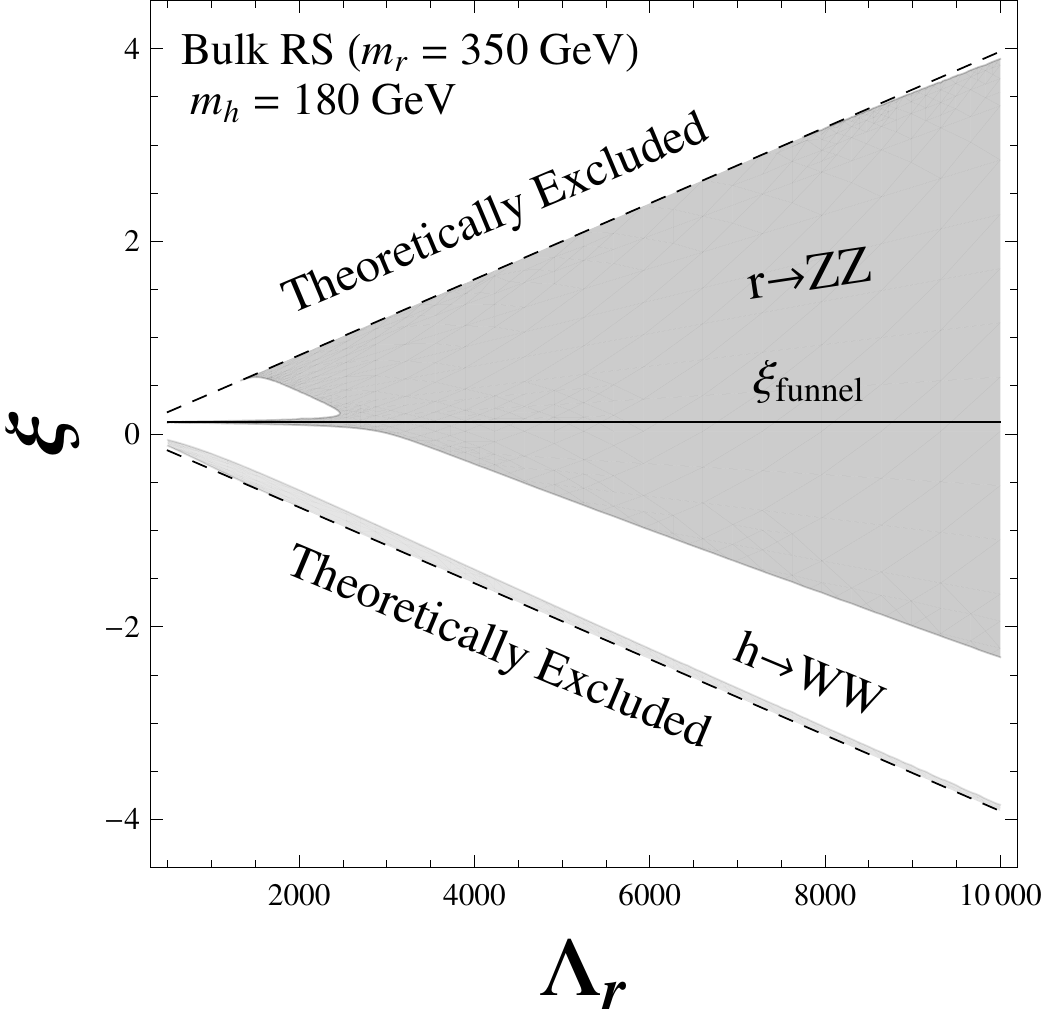}
\includegraphics[width=6.5cm]{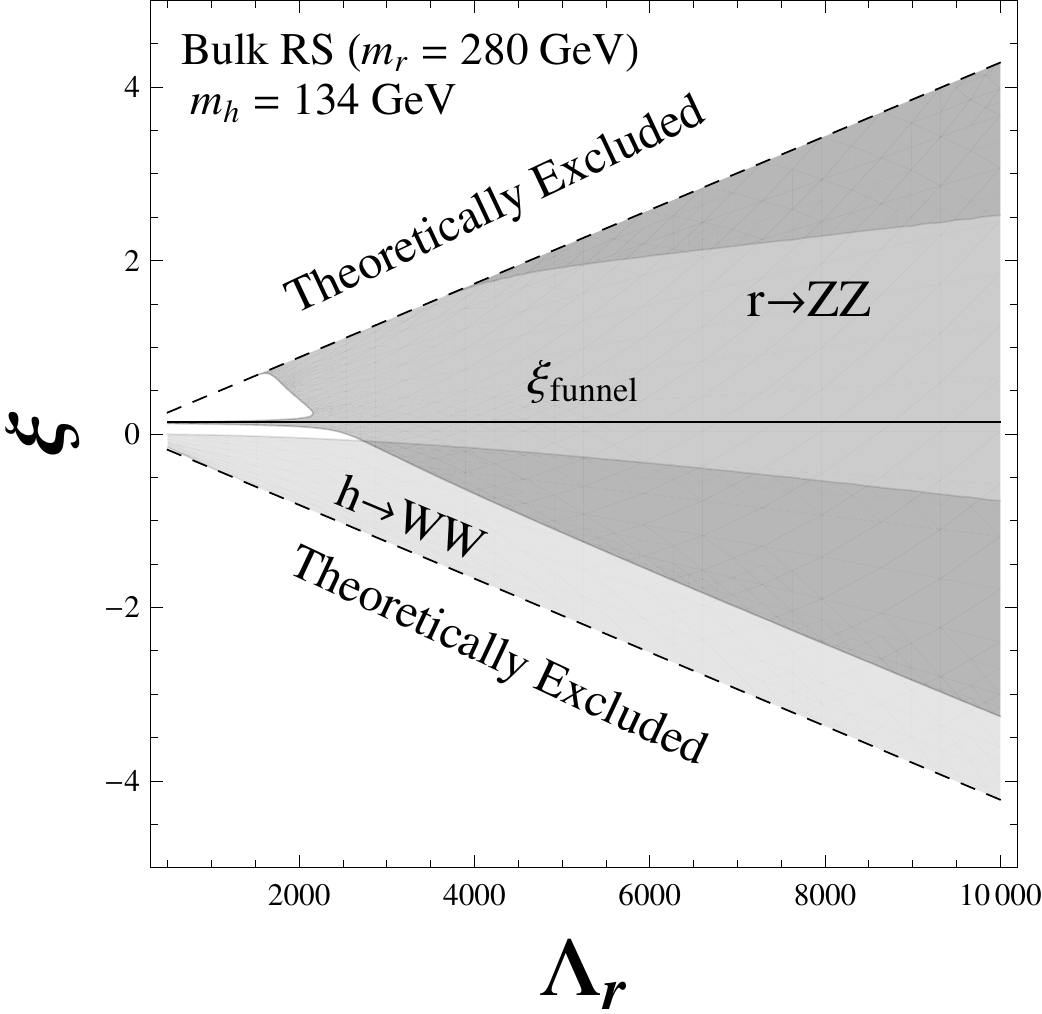}
\end{center}
\caption{Bounds on $\Lambda_r$ and $\xi$ coming from CMS combined data for two representative values of 
$m_r = 280, 350$ GeV and $m_h = 134, 180$ GeV, corresponding to a tightly constrained and a loosely constrained situations.
The white regions are excluded whereas the colored regions are allowed. 
The dotted lines correspond to the minimum and maximum 
values of $\xi$ theoretically allowed for a given value of $\Lambda_r$, $m_h$ and $m_r$. 
The region above (under) the maximum (minimum) values of $\xi$ is theoretically excluded. The white regions inside the
theoretically allowed $\xi$ range are experimentally excluded.
The light gray region is the one allowed by data for the Higgs and the dark gray for the radion. The intersections
of these two regions are the darker ones. The solid line are the values for $\xi_{funnel}$.
}
\label{Exclusion}
\end{figure}

In Fig.(\ref{Exclusion}) we show some of the exclusion regions in the $(\xi,\Lambda_r)$ plane 
for two extreme scenarios where the $WW$ and $ZZ$ constraints are strong ($m_r=350$ and $m_h=180$ GeV) and  
weak ($m_r=280$ and $m_h=134$ GeV). We apply the $WW$ channel constraints up to $200$ GeV and the $ZZ$ channel from $200$ to $600$ GeV.
The allowed region in the parameter space is somewhat enlarged for the less constrained situation but 
it remains challenging to have both a Higgs and a radion in the region 
where the SM Higgs has been excluded by the LHC.

\section{Conclusion}
In this paper we have studied the bounds on the parameters of the radion in a model of warped extra dimensions
obtained from CMS data on the Higgs boson searches presented at the end of 2011,
both in the vector boson and photon channels. 

We paid particular attention to the radion-Higgs  
mixing and showed that even for values of $\Lambda_r$ as low as $1$ TeV, the radion can still have a mass in the region 
where the Standard Model Higgs has been excluded, for a narrow range of values for the mixing parameter $\xi$. 
For a heavy radion this is only possible for a nonzero $\xi$. Hence, a low $\Lambda_r$ value with a radion within the LHC reach 
is only possible if a mixing between the Higgs boson and 
the radion is introduced. 
The $WW$ and $ZZ$ channels of the radion constraints the parameter $\xi$ to a narrow region close 
to $\xi_{funnel}$. 
In this case, the radion will be most likely detected in the $gg\rightarrow r \rightarrow hh$ channel, 
when $\xi=1/6$ or $gg\rightarrow r \rightarrow \gamma\gamma$ when $\xi=\xi_{funnel}$.

We also commented on the possibility of hiding the Higgs boson in this type of models, even for very large values of 
the radion scale. We find that the Higgs boson can still have a mass in the region 
where the Standard Model Higgs has been excluded for $\xi$ values close to its lower theoretical limit. Again the
mixing plays an essential role to evade the LHC constraints.

As the LHC collects more data in the near future, 
it will be possible to better scrutinize this class of models.

\section*{Acknowledgments}
We thank Christophe Grojean for reading the manuscript and providing useful comments, 
Gilad Perez for asking the question
about the possibility of hiding the Higgs, Heidi Rzehak for help with HDECAY and Pedro
Machado for help with the plots design.
RR would like to thank the TH division at CERN for its hospitality and the organizers 
of the Les Houches 2009 workshop on Physics at TeV Colliders,
where his interest in the radion started. 
HS would like to thank the TH division at CERN for its hospitality where part of this
work was done.
This work has been partially supported by the Sao Paulo state funding agency FAPESP (RR).
HS is supported by a CAPES Foundation(Ministry of Education of Brazil) Postdoc Fellowship
under the contract $n^{\underline{o}}$ BEX $1315/10-2$.

\end{document}